\def\preprint{0}                
\def\preprint{1}                
\def\comment#1{}
\if\preprint1
        \documentclass[usenatbib]{mn2e}
        \usepackage{times}
        \usepackage{graphicx}
        \usepackage{calc}

\else
        \documentstyle[astrop-bib,referee,times]{mn}
        \newcommand{\includegraphics}[1]{}
\fi

\def\oversim#1#2{\lower0.5pt\vbox{\baselineskip0pt \lineskip-0.5pt
     \ialign{$\mathsurround0pt #1\hfil##\hfil$\crcr#2\crcr\sim\crcr}}}
\def\aap{{\rm A\&A}}

\title[The low wind expansion velocity of metal-poor carbon stars 
in the Halo and the Sagittarius stream]{The low wind expansion 
velocity of metal-poor carbon stars in the Halo and the 
Sagittarius stream\thanks{Based on observations made at the 
James Clerk Maxwell Telescope under program M08AU01}\thanks{Based 
on observations made at the Very Large Telescope at Paranal 
Observatory under the  program 082.D-0836 }}
\author[E. Lagadec et al.]{Eric Lagadec$^{1}$ \thanks{E-mail:
eric.lagadec@manchester.ac.uk}
                Albert~A.~Zijlstra$^1$, Nicolas Mauron$^2$, Gary Fuller$^1$, Eric Josselin$^2$,\newauthor 
G.~C.~Sloan$^3$, A.~J.~E.~Riggs$^{4,5}$
\\
$^1$Jodrell Bank Center for Astrophysics
Alan Turing Building
School of Physics and Astronomy
The University of Manchester
Oxford Street
Manchester M13 9PL
UK \\
$^2$Goupe d'Astrophysique, UMR 5024 CNRS, Case CC72, 
Place Bataillon, 34095 Montpellier Cedex 5, France \\
$^3$Department of Astronomy, Cornell University, 
108 Space Sciences Building, Ithaca NY 14853-6801, USA\\ 
$^4$Astronomy Department, Yale University, 
260 Whitney Avenue, New Haven, CT 06511, USA\\
$^5$NSF REU Research Assistant, Department of Astronomy,
Cornell University, NY 14853-6801, USA\\
}
\pubyear{2008}

\begin{document}

\date{Accepted . Received}

\pagerange{\pageref{firstpage}--\pageref{lastpage}} \pubyear{2002}

\maketitle

\label{firstpage}

\begin{abstract}
We report the detection, from observations using the James 
Clerk Maxwell Telescope, of CO J\,$=$\,3$\rightarrow$\,2 
transition lines in six  carbon stars, selected as members 
of the Galactic Halo and having similar infrared colors.  Just one Halo star had been detected 
in CO before this work.  Infrared observations show that 
these stars are red (J-K $>$3), due to the presence of large dusty 
circumstellar envelopes.  Radiative transfer models 
indicates that these stars are losing mass with rather 
large dust mass-loss rates in the range 
1--3.3 $\times$$10^{-8}$M$_{\odot}$yr$^{-1}$, similar to 
what can be observed in the Galactic disc.  We show that 
two of these stars are effectively in the Halo, one is 
likely linked to the stream of the Sagittarius Dwarf 
Spheroidal galaxy (Sgr dSph), and the other three stars 
certainly belong to the thick disc.  The wind expansion 
velocities of the observed stars are low compared to 
carbon stars in the thin disc and are lower for the stars 
in the Halo and the Sgr dSph stream than in the thick disc.  
We discuss the possibility that the low expansion velocities 
result from the low metallicity of the Halo carbon stars. 
This implies that metal-poor carbon stars lose mass at a rate 
similar to metal-rich carbon stars, but with lower expansion 
velocities, as predicted by recent theoretical models.  This 
result implies that the current estimates of mass-loss rates 
from carbon stars in Local Group galaxies will have to be 
reconsidered.  
\end{abstract}



\section{Introduction}

Stars with initial masses in the range  0.8--8~M$_{\odot}$ 
end their life with a phase of catastrophic mass-loss.
During the asymptotic giant branch (AGB) phase, they develop 
a superwind leading to mass-loss rates up to 
10$^{-4}$M$_{\odot}$\,yr$^{-1}$. This superwind enriches the 
ISM with newly synthesized elements.

The mass-loss mechanism of AGB stars is likely due to 
pulsations from the star and radiation pressure on dust 
grains.  Shocks due to pulsation extend the atmosphere, so 
that the material ejected by the star becomes dense and 
cold enough for dust to form.  Due to its opacity, dust 
absorbs the radiation from the star and is driven away by 
radiation pressure, carrying the gas along through friction.
Theoretical models (Winters et al.\ 2000), show that the 
mass loss evolves from a pulsation driven regime 
characterized by a low mass-loss rate and a slow expansion 
velocity to a dust-driven regime with a high mass-loss rate 
and a high expansion velocity ($>$5km.s$^{-1}$).
 
Studying the effect of metallicity on the mass-loss 
process is important to understand the formation of dust 
around AGB stars in the early Universe.  At low metallicity, 
less seeds are present for dust formation, so one might 
expect dust formation to be less efficient and thus the 
mass-loss rates to be lower.

Theoretical work by Bowen \& Willson (1991) predicts that for 
metallicities below [Fe/H]$=-1$ dust-driven winds fail, and 
the wind must stay pulsation-driven.  However, observational 
evidence for any metallicity dependence is still very limited 
(Zijlstra 2004).  More recent observational (Groenewegen et 
al.\ 2007) and theoretical works (Wachter et al.\ 2008; 
Mattsson et al. 2008) indicate that the mass-loss rates from 
carbon stars in metal-poor environments are similar to our 
Galaxy.

To obtain the first observational evidence on mass-loss rates 
at low metallicity, we have carried out several surveys with 
the {\it Spitzer Space Telescope} of stars in nearby dwarf 
galaxies.  These show significant mass-loss rates down to 
Z=1/25 Z${_\odot}$ (Lagadec et al.\ 2007b; Matsuura et al.\ 
2007; Sloan et al.\ 2009), but only for carbon-rich stars. 
The current evidence indicates that oxygen-rich stars have 
lower mass-loss rates at lower metallicities.  For carbon 
stars, no evidence for a dependency of mass-loss rate on
metallicity has yet been uncovered.  Consequently, (Lagadec 
\& Zijlstra 2008) have proposed that the carbon-rich dust 
plays an important role in triggering the AGB superwind.

The main uncertainty arises from the unknown expansion 
velocity.  This parameter is needed to convert the density
distribution to a mass-loss rate.  The expansion velocity is 
also in itself a powerful tool.  Hydrodynamical simulations 
(Winters et al.\ 2000), have shown that radiation-driven 
winds have expansion velocity in excess of 5\,km\,s$^{-1}$, 
while pulsation-driven winds are slower.

\begin{figure}
\begin{center}
\includegraphics[width=9cm]{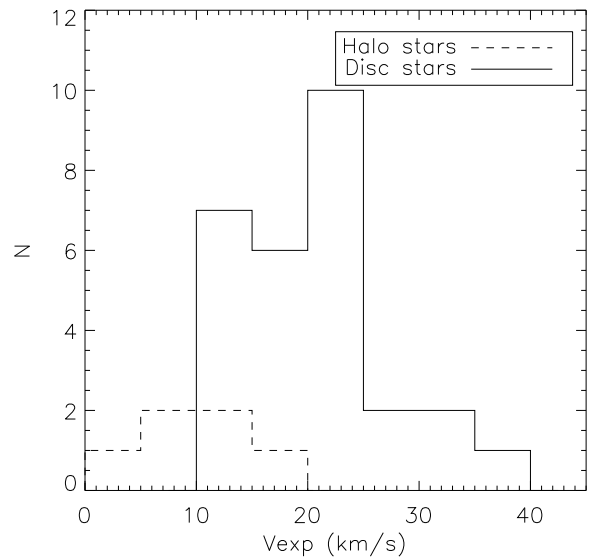}
\caption{\label{histo_vexp} Distribution of the expansion 
velocity for the observed Halo carbon stars compared 
with stars from the disc with similar J$-$K colors.}
\end{center}
\end{figure}

There is some evidence that expansion velocities are lower 
at low metallicity, from measurement of OH masers (Marshall 
et al.\ 2004).  However, OH masers are found only in 
oxygen-rich stars, which appear to have suppressed mass-loss
at low metallicities.  We lack equivalent measurements for
metal-poor carbon-rich stars, which do reach substantial 
mass-loss rates.  For these, the only available velocity 
tracer is CO.  Currently, extra-galactic stars are too 
distant for CO measurements.  However, there are a number of 
carbon stars in the Galactic Halo, which are believed to 
have similarly low metallicity.  One CO measurement exists
for a Galactic Halo star:  its expansion velocity has been 
estimated to be $\sim$3.2 km\,s$^{-1}$ through the 
$^{12}$CO $J=2 \rightarrow 1$ transition (Groenewegen et 
al.\ 1997).  This is much lower than typical expansion 
velocities, which are in the range 10-40 km\,s$^{-1}$ for stars with similar colours,
 and thus optical depths (Fig.\ref{histo_vexp}).

A large number of metal-poor carbon stars have recently 
been discovered in the Galactic Halo (Totten \& Irwin 1998; 
Mauron et al.\ 2004, 2005, 2007).  These stars may have been 
stripped from the Sagittarius Dwarf Spheroidal galaxy (Sgr
dSph), which has a metallicity of [Fe/H]$\sim-1.1$ (Van de 
Bergh 2000).  These stars are the closest metal-poor carbon 
stars known, and they are bright enough to be detected in CO 
using ground-based millimeter telescopes.

We have carried out observations of six Halo carbon stars in 
the CO J $= 3 \rightarrow 2$ transition.  Here, we report the 
results of these observations.

\section{Sample selection}

Mauron et al.\ (2004, 2005, 2007) and Mauron (2008) have 
discovered $\sim$ 100 carbon stars in the Galactic Halo, 
adding to the sample of $\sim$ 50 described by Totten \& 
Irwin (1998).  All of these stars are spectroscopically 
confirmed carbon stars.  Only the brightest can be detected 
in the sub-millimeter range.  We selected the six stars with 
the highest IRAS 12$\mu$m flux observable with the James 
Clerk Maxwell Telescope (JCMT, Mauna Kea, Hawaii).  The 
emission from an AGB star at 12$\mu$m is due to thermal 
emission from the dust in the envelope.  Thus one expects 
the stars with the largest 12$\mu$m flux to be the brightest 
in CO. All the observed stars have 3$<J-K<$4 and have thus a similar dust optical depth and a large circumstellar dusty envelope.

Table\,\ref{targets} describes the observational properties 
of the selected stars.  One of these stars, IRAS 12560+1656 
(Beichman et al.\ 1990), is the only Halo carbon star that 
had previously been detected in CO J$=2 \rightarrow 1$ (by
Groenewegen et al.\ 1997).  

\begin{table*}
\caption[]{\label{targets} Observed Halo stars targets:
names, adopted coordinates, photometry and distance.
J, H, K$_s$ are taken from 2MASS. f$_{8.59}$, f$_{13.04}$ and f$_{17.65}$ are the VISIR/VLT flux at 8.59, 
13.04 and 17.65 $\mu$m respectively. f$_{12}$ is the IRAS 12$\mu$m flux. Z is the distance above the Galactic plane.}
\begin{center}
\begin{tabular}{llllllllllllllll}
\hline
IRAS name& Other name& RA & Dec  
& K & J$-$K&f$_{8.59}$&f$_{13.04}$&f$_{17.65}$&f$_{12}$& [12]& Z\\
  && \multicolumn{2}{c}{(J2000)}& mag &mag &Jy & Jy&Jy&Jy&mag& kpc\\
\hline

IRAS 04188$+$0122&CGS 6075 &04 21 27.25&$+$01 29 13.4  & 6.420 &  3.280 &4.99 &2.67 & 1.64  & 3.37& 2.31  & -2.63\\
IRAS 08427$+$0338&CGS 6306 &08 45 22.27&$+$03 27 11.2  & 6.255 &  3.410 &- &- &-   & 6.50& 1.60  &  2.03\\
IRAS 11308$-$1020&CGS 3052 &11 33 24.57&$-$10 36 58.6  & 4.568 &  3.830 &- &- &-   &57.37& -0.77 &  1.41\\
IRAS 16339$-$0317&CGS 3716 &16 36 31.70&$-$03 23 37.5  & 6.098 &  3.906 &12.49 &10.83 &5.60   &14.57& 0.72  &  1.73\\
IRAS 12560$+$1656&CGS 6500 &12 58 33.50&$+$16 40 12.0  & 7.820 &  3.480 &- &- &-   & 0.77& 3.91  &  8.96\\
IRAS 18120$+$4530&         &18 13 29.6 &$+$45 31 17.0  & 6.710 &  3.809 &- & -& -  & 7.86& 1.39  &  2.20\\

\hline \\
\end{tabular}
\end{center}
\end{table*}


\section{Observations and data reduction}

\begin{figure*}
\begin{center}
\includegraphics[width=18cm]{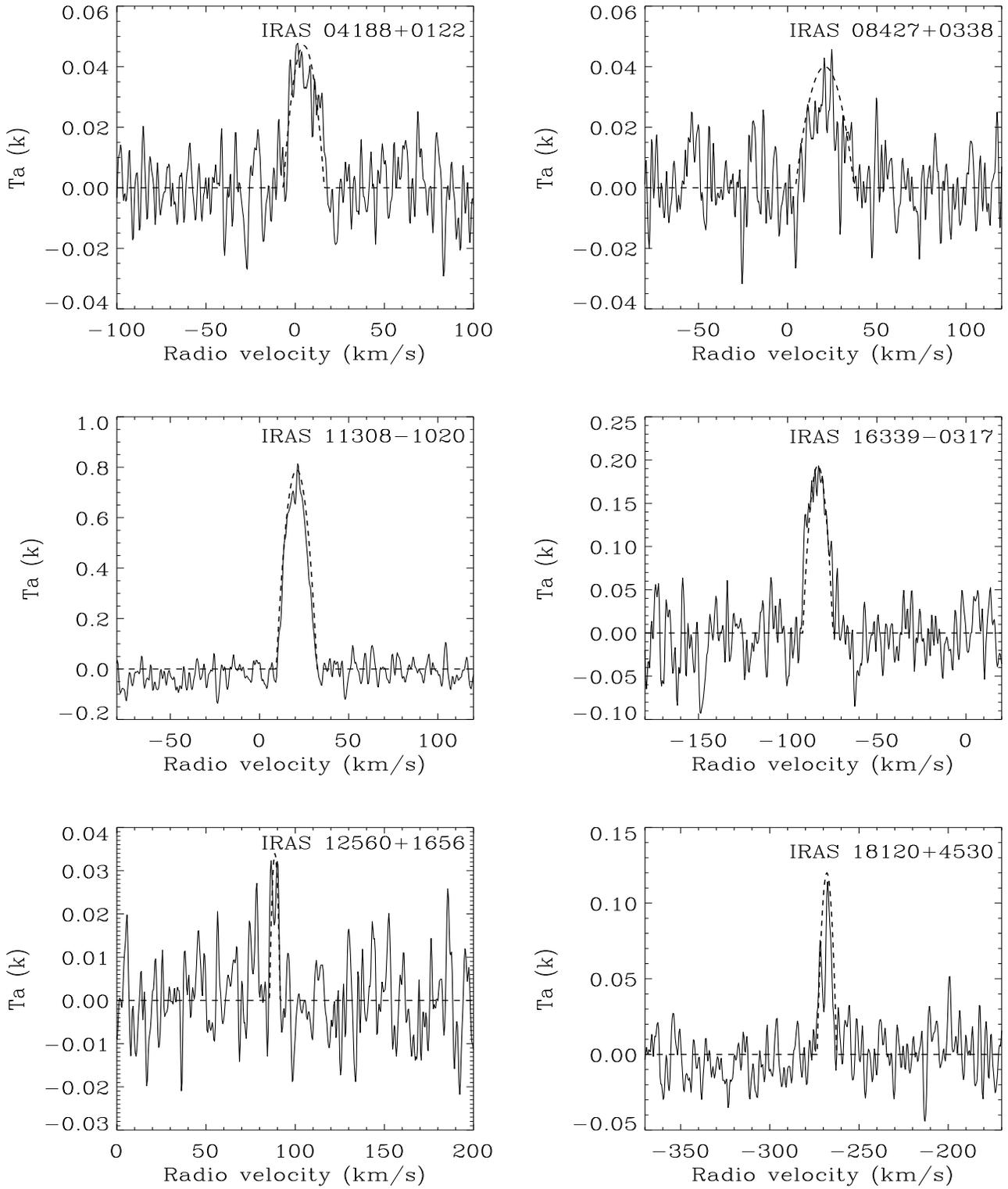}
\caption{\label{co_obs}JCMT CO 3--2 line of the 6 observed stars. 
The dashed line represent the best model fit to
the data. The velocities in abscissa are V$_{LSR}$}
\end{center}
\end{figure*}

We observed the CO J\,$=$\,3$\rightarrow$\,2 (345 GHz) line emission 
from the six selected carbon stars from the JCMT using the 
heterodyne focal-plane array receiver (HARP; Smith et al.\ 
2008).  Observations were carried with a bandwidth of 1 GHz 
and a frequency resolution of 0.977 MHz, giving a velocity 
coverage of $\pm$400 km\,s$^{-1}$.  To increase the 
signal-to-noise ratio of our observations, we used a 30'' 
beam switch to keep one detector on the source at all times.  
We used the Starlink software to reduce the resulting data.  
Fig. \ref{co_obs} shows the six CO 3--2 lines we detected.

We obtained mid-infrared photometric observations of two 
stars from our sample using VISIR (Lagage et al.\ 2004) on 
the VLT.  The observations were taken using four filters 
centred at 8.59$\mu$m (PAH1, $\Delta$$\lambda$=0.42$\mu$m),
11.25$\mu$m (PAH2, $\Delta$$\lambda$=0.59$\mu$m), 
13.04$\mu$m (NeII\,2, $\Delta$$\lambda$=0.22$\mu$m)
and 17.65$\mu$m (Q1, $\Delta$$\lambda$=0.83$\mu$m).  We
used the standard mid-infrared chop-and-nod technique to 
remove the background emission from the telescope and sky.
Standard stars were observed just after each observation for 
flux calibration.  The data were reduced using the IDL 
routines developed and described by Lagadec et al.\ (2008).

\begin{table*}
\caption[]{\label{line_values} CO J=3--2 lines properties 
of the observed Halo stars. Typical errors on the expansion velocities estimates are 0.1 km.s$^{-1}$ for the stars with expansion velocities above 10km.s$^{-1}$  and 0.2s$^{-1}$ for the other ones}
\begin{center}
\begin{tabular}{llllllllllllllll}
\hline

 Adopted name & peak & V$_{LSR}$ & $V_{\rm exp}$ & FWHM & flux & rms& Integration time   \\
  & K& km.s$^{-1}$& km.s$^{-1}$& km.s$^{-1}$ &K.    km.s$^{-1}$& K & mn\\
\hline
           
           IRAS 04188+0122  & 0.04     &   4.8  &  11.5  & 17.3  &  0.77  &  0.009&616\\
           IRAS 08427+0338  & 0.03     &  20.9  &  16.5  & 17.4  &  0.55  &  0.010&45\\
           IRAS 11308-1020  & 0.79     &  20.8  &  11.5  & 12.9  & 10.86  &  0.049&90\\
           IRAS 16339-0317  & 0.19     & -83.1  &   8.5  & 12.6  &  2.59  &  0.030&182\\
           IRAS 12560+1656  & 0.03     &  88.6  &   3.   &  4.5  &  0.14  &  0.008&453\\
           IRAS 18120+4530  & 0.08     &-268.0  &   6.5  &  8.3  &  0.72  &  0.019&227\\
           
\hline \\
\end{tabular}
\end{center}
\end{table*}

%
%

%

 \section{CO line properties}
Fig.\ \ref{co_obs} shows that the six lines we detected have 
parabolic profiles, which arise when the envelopes of the 
observed stars are spherical, unresolved by the telescope, 
have a constant expansion velocity $V$$_{\rm exp}$,
and the CO 3--2 line is optically thick (Knapp \& Morris 1985). 
This can be described as:

\begin{equation}
T^*_A(V)=T^*_A(\rm peak)\left[1-\left(\frac{V-V_c}{V_0}\right)^2\right],
\end{equation}

\noindent where $T^*_A(\rm peak)$ is the peak emission 
temperature, $V_0$ is the expansion velocity of the envelope, 
and V$_C$ the velocity of the line.  We fitted this function 
to the observed data with a least-squares technique.  
Table\,\ref{line_values} lists the resulting properties of 
the observed lines. The CO lines detected for IRAS\,12560+1656 is rather weak, but can be considered as a real detection
as it was already detected in another transition by Groenewegen et al. (1997).

\begin{figure}
\begin{center}
\includegraphics[width=9cm]{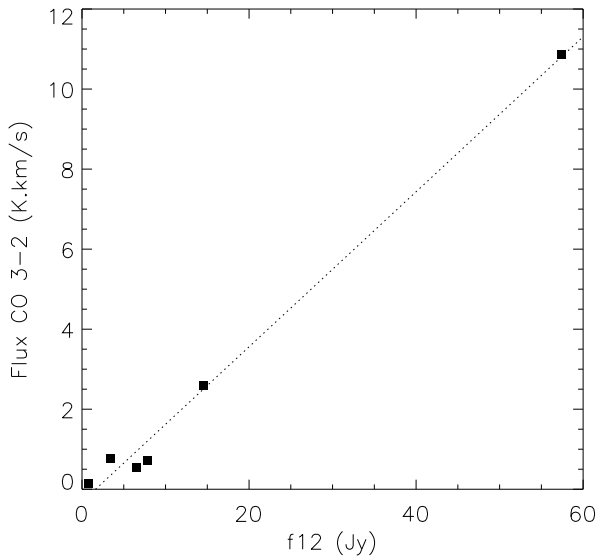}
\caption{\label{f12} The linear correlation between the IRAS 
12$\mu$m flux and the observed CO J=3--2 flux.}
\end{center}
\end{figure}

Fig.\,\ref{histo_vexp} shows the distribution of expansion 
velocities for our sample.  As expected, the flux measured in
the CO 3--2 line correlates strongly with the IRAS 12$\mu$m 
flux (Fig.\,\ref{f12}). The best linear fit to the data is 
obtained for:

\begin{equation}
F_{\rm CO}=0.194 \pm 0.009 \times f_{12} -0.314 \pm 0.214
\end{equation}


\section{Dust mass-loss rates}


Two techniques are usually used to estimate dust mass-loss 
rates from AGB stars:  by measuring the near- and mid-infrared 
flux of the stars (Whitelock et al.\ 1994, 2006; Lagadec et 
al.\ 2008) or by using radiative transfer models.  We apply
both methods here.  We estimated the mass-loss rates by 
measuring the K$_s$$-$[12] colour, where K$_s$ is the 2MASS 
2.16$\mu$m magnitude and [12] the IRAS 12$\mu$m magnitude, 
according to the relation from Whitelock et al. (2006):

\begin{eqnarray}
\label{mass-loss}
\nonumber \log(\dot{M}_{\rm total}) &&=-7.668+0.7305(K-[12])\\
\nonumber  &&-5.398 \times10^{-2}(K-[12])^2  \\
  &&+1.343\times10^{-3}(K-[12])^3,
\end{eqnarray}

\noindent where [12] is the IRAS 12$\mu$m magnitude assuming 
a zero-magnitude flux of 28.3 Jy.
 
\begin{table*} 
\begin{center} 
\caption[]{\label{per_ml} Dust mass-loss rates for the 
observed stars and parameters obtained from our DUSTY models. 
$\dot{M}_{\rm col}$ and $\dot{M}_{\rm dusty}$ are the dust 
mass-loss rates  (in M$_{\odot}$yr$^{-1}$) from Lagadec et 
al.\ (2008) and our DUSTY models, respectively. The distance 
we adopted is the average between the distances D$_1$ and D$_2$ 
described in Section \ref{distances}} 
 
\begin{center} 
\begin{tabular}{cccccccccccccccccccrlllllll} 
\hline 
Target& Luminosity &$\dot{M}_{\rm DUSTY}$ &T$_{eff}$& T$_{\rm in}$& SiC/AMC& 
      $\tau$ (0.55$\mu$m)& $\dot{M}_{\rm col}$& Distance  \\ 
      & (L$_\odot)$   &  (10$^{-8}$M$_{\odot}$yr$^{-1}$) &(K)&(K)&&& 
              (10$^{-8}$M$_{\odot}$yr$^{-1}$) & kpc \\ 
\hline 
IRAS 04188+0122 & 8221   &  1.0   & 2800 &  1200  & 0.1   & 10.   &  1.6 & 6.2   \\ 
IRAS 08427+0338 & 10124  &  1.8   & 2800 &  1200  & 0.    & 10.   &  2.5 & 5.6  \\ 
IRAS 11308-1020 & 16793  &  2.9   & 2800 &  1200  & 0.    & 10.   &  3.9 & 2.4  \\ 
IRAS 16339-0317 & 6976   &  1.8   & 2800 &  1200  & 0.    & 17.8  &  4.0 & 4.8  \\  
IRAS 12560+1656 & 7563   &  0.4   & 2800 &  1200  & 0.    & 10.   &  1.4 & 11.4  \\ 
IRAS 18120+4530 & 13391  &  1.4   & 2800 &  1200  & 0.1   & 10.   &  3.9 & 6.5  \\ 
 
 \hline \\ 
\end{tabular} 
\end{center} 
\end{center} 
\end{table*} 
 
We obtained an alternative estimate of the dust mass-loss 
rates using a radiative transfer model.  We used the 
radiative transfer code DUSTY (Ivezi\'c \& Elitzur 1997). 
This code solves the 1-dimensional problem of radiation 
transport in a dusty environment.  For all of our models, we 
assume that the irradiation comes from a point source (the 
central star) at the centre of a spherical dusty envelope. 
The circumstellar envelope is filled with material from a 
radiatively driven wind.  All of the stars are carbon-rich,
and the dust consists of amorphous carbon.  We did
not add SiC to our model, except for IRAS\,04188+0122 and IRAS\,16339-0317, for 
which a fit was impossible without SiC.  Optical properties 
for these dust grains are taken from Hanner (1988) and 
Pegouri\'e (1988) for amorphous carbon and SiC, respectively. 
The grain size distribution is taken as a typical MRN
distribution, with a grain size $a$ varying from 0.0005 to 
0.25$\mu$m distributed according to a power law with 
n($a$)$\propto$$a^{-q}$ with $q$=3.5 (Mathis et al.\ 1977).  
The outer radius of the dust shell was set to 10$^3$ times 
the inner radius; this parameter has a negligible effect on 
our models.
 
To model the emission from the central star, we used a 
hydrostatic model including molecular opacities (Loidl 
et al.\ 2001; Groenewegen et al.\ 2007).  Our aim was to fit 
the spectral energy distribution, defined by the 2MASS 
photometry (J, H and K), our VISIR mid-infrared data, and the 
IRAS photometry (at 12 and 25$\mu$m) to estimate the dust 
mass-loss rates.  We fixed the dust temperature at inner 
radius (1220K), the mass ratio of SiC to amorphous (0 $\%$)
carbon dust and the effective temperature of the central star,
T$_{\rm eff}$ (2800K), unless a satisfactory fit could not be
obtained with these parameters.

DUSTY gives total (gas+dust) mass-loss rates assuming a 
gas-to-dust ratio of 200. Estimating the gas-to-dust ratio of these stars is beyond the scope of this paper 
and will be the subject of a forthcoming paper using CO observations of these stars in other transitions and Spitzer spectra.
We thus prefer to estimate dust mass-loss rates.
  The dust mass-loss rate can be 
obtained by dividing by the gas-to-dust mass ratio, assuming 
that the gas and dust expansion velocities are the same. 
Here we prefer to talk about dust mass-loss rates as inputs from our models
as we are fitting infrared colours, and thus dust  emission.
 The 
expansion velocity is an output for the DUSTY models, and we 
scaled our results using the expansion velocities we measured.
Table\,\ref{per_ml} gives the results of the fits. The IRAS 12$\mu$m flux is
an outlier for IRAS 16339$-$0317. This is certainly due to variability or to the fact that the aperture of the IRAS satellite is larger than the VISIR one.


\begin{figure*}
\begin{center}
\includegraphics[width=18cm]{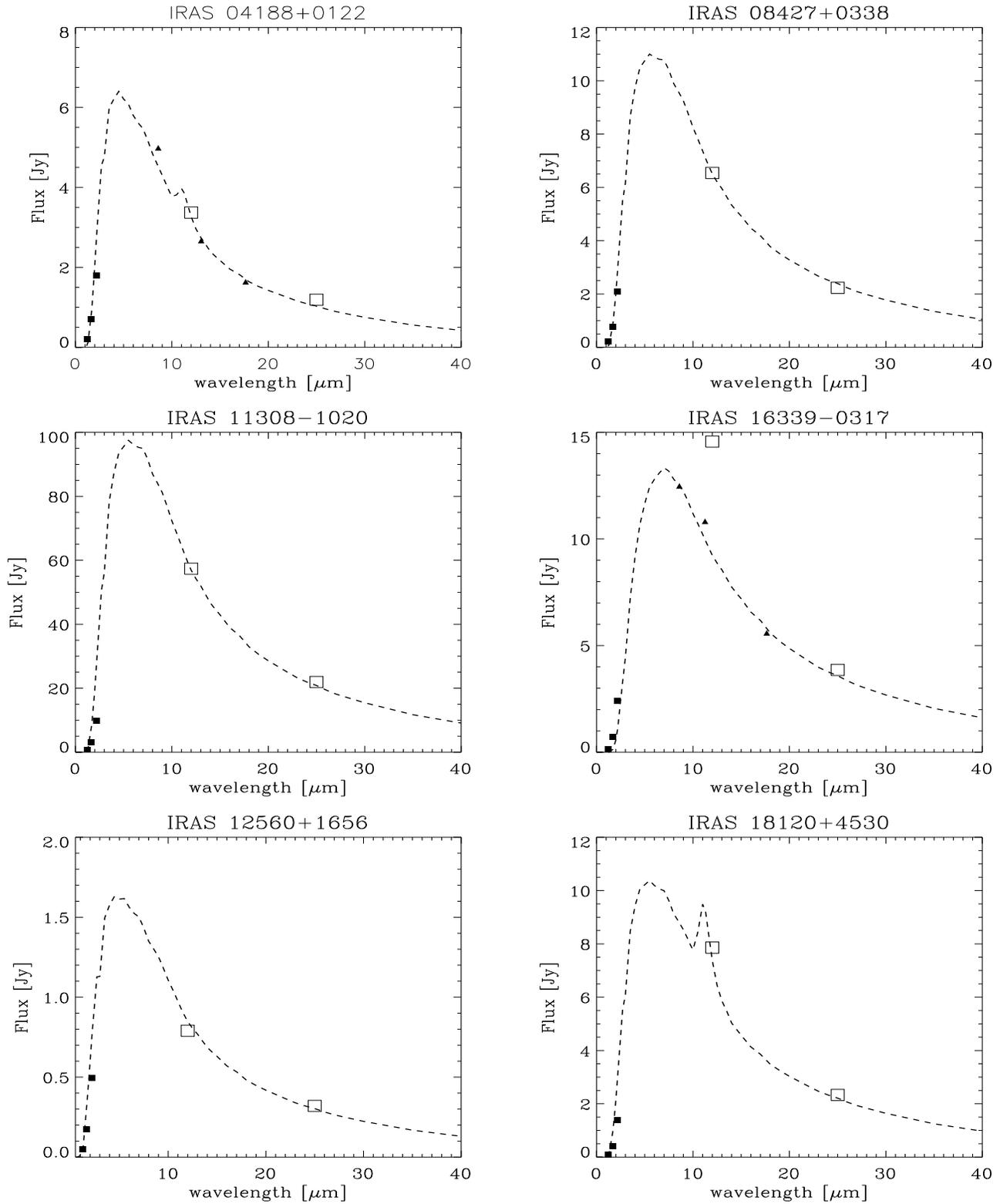}
\caption{\label{dusty}Radiative transfer DUSTY models of the 
spectral energy distributions of the observed stars.  The 
dashed lines represent our best models, and the symbols are 
flux from 2MASS, IRAS (12 and 25$\mu$m) and our VISIR 
observations.  Filled squares, open squares and triangles 
represent 2MASS, IRAS and VISIR/VLT data, respectively.}
\end{center}
\end{figure*}

\section{Discussion}

\subsection{Mass loss and evolution}

The detection of resolved CO lines in our six targets 
indicate that all of these stars are losing mass.  Our VISIR 
observations, as well as IRAS and 2MASS observations, show 
that these stars are very red, which indicates the presence 
of circumstellar dust.  To study the impact of the mass loss 
on the evolution of the studied stars, we compared the 
measured mass-loss rates with nuclear reaction rates and the 
classical single-scattering limit (see e.g., van Loon et 
al.\ 1999 and Lagadec et al.\ 2008).

\begin{figure}
\begin{center}
\includegraphics[width=9cm]{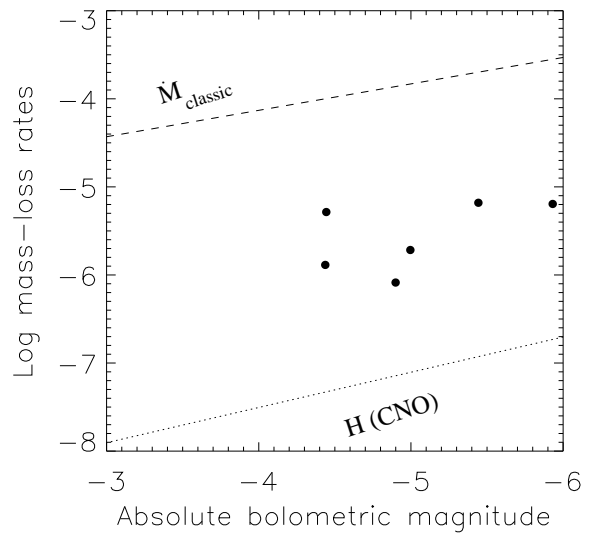}
\caption{\label{mbolmlr} Total (gas+dust) mass-loss rate as 
a function of the absolute bolometric magnitude.}
\end{center}
\end{figure}

Fig.\ref{mbolmlr} shows the total (gas+dust) mass-loss rate 
as a function of the absolute magnitude.  The absolute 
magnitude is taken from our DUSTY models. The total mass-loss 
rate is also an output from our DUSTY models, assuming a 
typical gas-to-dust mass ratio of 200.  The dotted line shows 
the rate at which mass is consumed by nuclear burning and the 
dashed line the classical single-scattering limit as 
described by Lagadec et al.\ (2008).  This shows that no 
multiple scattering is needed to explain the observed dust 
mass-loss rates.  The mass-loss rates we measure are also 
well above the nuclear burning rate line, indicating that the 
evolution of these stars will be governed by the mass-loss 
process.  Similar conclusions were drawn from observations 
of metal-poor AGB stars in the LMC, Fornax and Sgr dSph (van 
Loon et al.\ 1999, Lagadec et al.\ 2008).

\subsection{Distances}
\label{distances}

\begin{table*}
\caption[]{\label{gal}Galactic location of our observed stars. 
D$_1$, D$_2$ and D$_3$ are the distance to the sun using
methods described by Sloan et al.\ (2008), Mauron et al.\ (2008) 
and the period-luminosity relation respectively.  This last
method was not applied for three stars without known periods 
from the Northern Sky Variability Survey (NSVS, Wo\`zniak et al. 2004).}
\begin{center}
\begin{tabular}{llllllllllllllll}
\hline
Adopted name& l & b  & Period&D$_1$&D$_2$&D$_3$&$<$D$>$\\
  &&&days&kpc&kpc &kpc &kpc\\
\hline

IRAS 04188+0122 & 192.1775&-31.9867&359 &7.3 &6.0 &6.4& 6.5$\pm$0.6\\
IRAS 08427+0338 & 223.4859&+26.8173&288 &6.6 &5.3 &5.1& 5.5$\pm$0.8\\
IRAS 11308-1020 & 273.6969&+47.7772&-   &2.8 &2.1 &-  & 2.5$\pm$0.5\\
IRAS 16339-0317 & 012.7346&+27.7944&-   &5.6 &4.2 &-  & 4.9$\pm$1.0\\
IRAS 12560+1656 & 312.2528&+79.4127&-   &13.3&10.7&-  & 12.0$\pm$1.9\\
IRAS 18120+4530 & 073.0530&+25.3482&408 &7.6 &5.7 &6.7& 6.7$\pm$0.9\\

\hline \\
\end{tabular}
\end{center}
\end{table*}


\begin{figure}
\begin{center}
\includegraphics[width=9cm]{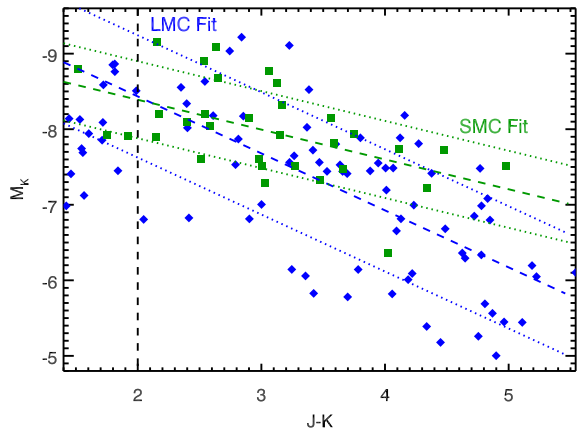}
\caption{\label{fig_dist} Absolute K magnitude (M$_K$) of carbon stars as a fonction of their J-K infrared colours. 
Diamonds and squares represent LMC and SMC carbon stars respectively. The vertical dashed line represent the limit 
under which relation Eq.\ref{dist_lmc} is no longer valid.}
\end{center}
\end{figure}

Most of the methods to determine mass-loss rates rely on an
accurate distance to the observed stars.  To determine 
these distances, we applied three methods.  Two are  based on 
near-infrared colours; the third is the period-luminosity 
relationship.


The first method uses the  infrared colours of the observed stars, 
using the relation between M$_K$ and J$-$K determined by
Sloan et al. (2008).  They found  from a sample of carbon
stars in the Small Magellanic Cloud (SMC) that:
 
\begin{equation} 
\label{dist_smc}
M_K = -9.18 + 0.395(J-K) 
\end{equation}  

Mauron et al.\ (2008) compared the near-infrared 2MASS 
photometry of stars in the Halo and the Large Magellanic
Cloud (LMC) and found a different relationship from the one 
above between J$-$K and M$_K$.  Their relation gives fainter 
M$_K$ values at a given J$-$K colour than that of Sloan et 
al., with the difference increasing with redder colours.  We 
have recalibrated the relation for the LMC, using the samples 
of stars confirmed to be carbon-rich with the Infrared 
Spectrograph on {\it Spitzer} by Zijlstra et al.\ (2006), 
Buchanan et al.\ (2006), Leisenring et al.\ (2008), and Sloan 
et al.\ (2008).  We find that:

\begin{equation} 
\label{dist_lmc}
M_K = -9.94 + 0.754(J-K),
\end{equation}  

\noindent for J$-$K colours greater than 2.0. 
 This 
calibration of the J$-$K relation is our second method.
Fig.\, \ref{fig_dist} compares the SMC and LMC calibrations of the relation
between M$_K$ and J$-$K.

The scatter in the LMC sample is 0.81 magnitudes about the
fitted line, compared to 0.51 magnitudes for the SMC sample.
The two fitted lines yield nearly identical distances at
J$-$K = 2, but as the colour grows redder, the samples
diverge from each other.  At J$-$K = 5, the difference in
$M_K$ is a full magnitude.  For the colours in our sample,
the two methods yield results differening by 0.41 to 0.64
magnitudes, comparable to the spread in the SMC sample.
The different slopes in the two samples may result from
their different metallicities, but that is only speculation
on our part.

The third method utilises the period-luminosity relation for 
carbon Miras described by Feast et al.\ (2006): 
 
\begin{equation} 
M_{\rm bol}=-2.54  \log P+2.06, 
\end{equation} 
With an uncertainty of 0.24 magnitudes. 
We derived the bolometric magnitudes using the equation for
bolometric correction derived by Whitelock et al.\ (2006), 
after converting all of the photometry to the SAAO system as 
described by Lagadec et al.\ (2008):
 
\begin{eqnarray} 
\nonumber {\rm BC_K} & = & +\, 0.972 + 2.9292\times(J-K) 
  -1.1144\times(J-K)^2 \\ 
 & & +0.1595\times(J-K)^3 -9.5689\,10^3(J-K)^4 
\end{eqnarray} 
\noindent  

Table \ref{gal} presents the obtained distances, D$_1$, D$_2$ 
and D$_3$ respectively.  For those stars without periods, the 
final estimated distance is the average of D$_1$ and D$_2$.  
For those stars with periods, the D$_1$ and D$_2$ values
bracket D$_3$ in two of the three cases.  Consequently, we 
first averaged D$_1$ and D$_2$, then averaged the result 
with D$_3$ to arrive at our final estimate of the distance.
The uncertainties in the final estimated distances  are the the standard deviation of the individual
distances. 

\subsection{Galactic location of the stars}
\label{location}

Knowing the distance to the stars and their $V_{\rm lsr}$ 
allows us to study their location in the Galaxy.  All six
stars examined here have been classified previously as 
members of the Halo, based on their distances from the 
Galactic plane.  Our CO observations give us velocity 
information for these stars, which we can compare to Galactic 
rotation models.  Table \ref{gal} lists the Galactic 
coordinates l and b.

\begin{figure*}
\begin{center}
\includegraphics[width=13cm,angle=-90]{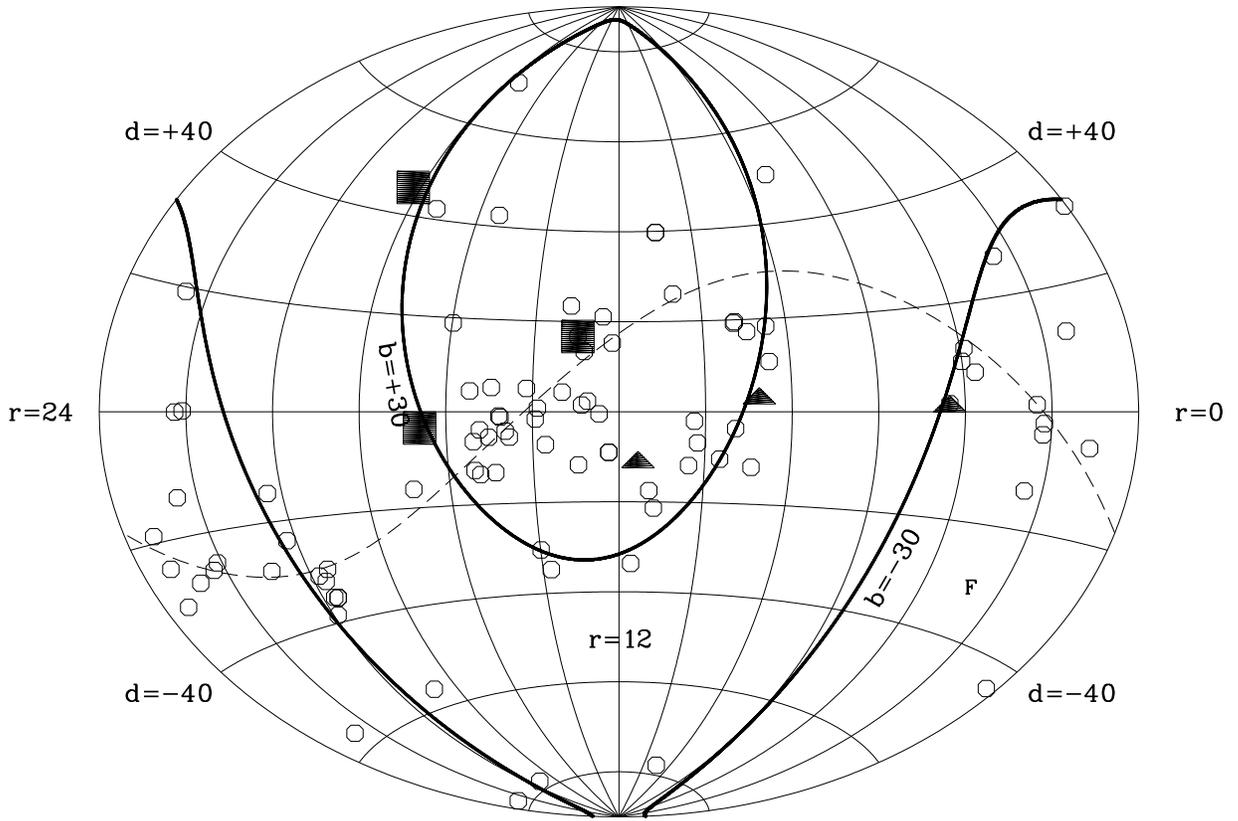}
\caption{\label{aitoff} The location of the six carbon
stars on an Aitoff projection of the sky.  Open circles 
represent Halo carbon stars from Mauron et al.\ (2004, 
2005, 2007) and Totten \& Irwin (1998) with J$-$K $>$1.2 
and K$>$ 6 to eliminate carbon stars in the disc.  Filled 
symbols represent the six stars in our sample.}
\end{center}
\end{figure*}

Fig.\ref{aitoff} shows the location of our six carbon stars
on an Aitoff projection.  The dashed line schematically 
represents the Sgr dSph orbit (Ibata et al. 2001).  One of 
our stars, IRAS\,12560+1656, lies very close to this orbit.  Its 
$V_{\rm lsr}$ is in the range of the observed $V_{\rm lsr}$ 
for stars in the Sgr dSph stream. IRAS\,12560+1656 very likely belongs 
to this stream.  Our observations are thus certainly the 
first CO observations of an extragalactic AGB star.  Two 
other stars, IRAS\,16339-0317 and IRAS\,18120+4530, have large negative 
$V_{\rm lsr}$, fully consistent with membership in the 
Halo.  Finally, IRAS\,04188+0122, IRAS\,08427+0338, and IRAS\,11308-1020 have a distance, 
location and $V_{\rm lsr}$ consistent with membership in the 
thick disc.  These three last stars thus have a metallicity 
between that of the thin Galactic disc and the Galactic 
Halo, the average metallicity of the thin disc being $\sim$-0.17 while the one of the thick disc is 
$\sim$-0.48 (Soubiran et al., 2003).
Our sample thus contains three metal-poor AGB stars and three 
AGB stars with intermediate metallicity.

\subsection{Low expansion velocities in the Halo}

\label{lowvexp}
The present observations allow us to directly measure the 
expansion velocity for a sample of carbon stars in the Halo.
The stars we observed are quite red (3$<$J$-$K$<$4), and have 
substantial circumstellar dusty envelopes responsible 
for the observed reddening.  Our measured expansion 
velocities are in the range 3--16.5 km\,s$^{-1}$.  The 
expansion velocity of carbon stars increases during the 
evolution on the AGB (Scho\"ier, 2007), i.e.\ when the dusty 
envelope becomes optically thicker.  To compare the expansion 
velocities we measured in Halo carbon stars with carbon stars
in the disc, we took a sample of carbon stars in the disc 
with colours similar to our sample.  We selected all of the
carbon stars with 3$<$J$-$K$<$4 in the extensive catalogue of 
CO observations of evolved stars by Loup et al.\ (1993).  
Fig.\ref{histo_vexp} compares the distribution of expansion 
velocities in the two samples.  The Halo carbon stars clearly
have a lower mean expansion velocity.

The three stars in the Halo and the Sgr dSph stream have 
$V_{\rm exp}$ in the range 3--8.5 km\,s$^{-1}$, while the 
three stars associated with the thick disc have velocities 
ranging from 11.5 to 16.5 km\,s$^{-1}$.  The latter range is
at the low end of expansion velocities for AGB stars with
similar near-infrared colours in the thin disc.

\subsection{Origin of the low expansion velocities}

Section \ref{location} and \ref{lowvexp} have shown that the 
three stars we observed in the Halo and Sgr dSph stream have 
low expansion velocities.  The stars we observed in the thick 
disc have expansion velocities intermediate between those 
typically observed in the Halo and the thin disc.  This 
difference could arise from differences in metallicity, with
the more metal-poor carbon stars having the slower winds.

Mattsson et al.\ (2008) and Wachter et al.\ (2008) have 
recently conducted theoretical investigations of the winds 
from metal-poor carbon stars.  Both studies show that 
metal-poor carbon stars can develop high mass loss rates, 
leading to the formation of a large dusty envelope, in 
agreement with spectroscopic observations from {\it Spitzer} 
of AGB stars in metal-poor galaxies (Zijlstra et al.\ 2006; 
Sloan et al.\ 2006;  Groenewegen et al.\ 2007; Lagadec et 
al.\ 2007; Matsuura et al.\ 2007; Leisenring et al.\ 2008;
Lagadec et al.\ 2009; Sloan et al.\ 2009).  

Wachter et al.\ (2008) predict that the outflow velocities 
from carbon stars should be lower in metal-poor environments,
because of the lower gas-to-dust mass ratio and because the 
formation of less dust leads to less efficient acceleration 
of the wind outside of the sonic region.  This interpretation
is consistent with our interpretation that the low expansion 
velocities we have observed in the Halo are due to their low 
metallicities.

Hydrodynamical models (Winters et al.\ 2000, Wachter et al.\ 
2008) distinguish two types of models.  Model A applies to
cases where the radiation pressure on dust is efficient. 
Mass-loss rates can exceed 10$^{-7}$M$_{\odot}$\,yr$^{-1}$, 
and expansion velocities can climb above $\sim$5 km\,$s^{-1}$.
Model B applies to cases where pulsations drive the mass loss.
In these cases, the mass-loss rates and expansion velocities
are smaller.  The mass loss occurs in a two-step process, with
stars first losing mass due to pulsations, followed by 
acceleration due to radiation pressure on the dust grains. The 
high mass-loss rate and low expansion velocity of TI~32 does not 
fit either of these models, possibly because metal-poor carbon
stars can develop strong mass-loss from pulsation alone. 


\section{Conclusions and perspectives}

We have detected the CO J\,$=$\,3$\rightarrow$\,2 in six carbon stars 
selected as Halo stars.  Only one carbon star had been 
detected in CO previously.  Comparison of the infrared 
observations and radiative transfer models indicates that 
these stars are losing mass and producing dust.  Their 
mass-loss rates are larger than their nuclear burning rates, 
so their final evolution will be driven by this mass-loss 
phenomenon.

We show that three of the observed stars are certainly 
members of the thick disc, while one is in the Sgr dSph stream 
and two are in the Halo.  The CO observation of the Sgr dSph 
stream star are thus the first identified millimetre 
observations of an extragalactic AGB star.  The expansion 
velocity we determined from our CO observations are lower 
than those of carbon stars in the thin disc with similar 
near-infrared colours.  The observed carbon stars with the 
lowest expansion velocities are Halo or Sgr dSph stream carbon 
stars.  There is a strong indication that the expansion winds 
are lower in metal-poor environments, which agrees with recent 
theoretical models (Wachter et al.\ 2008). 

So far, the effect of metallicity on the mass-loss from 
carbon-rich AGB stars has studied primarily from infrared 
observations of AGB stars in Local Group galaxies, mostly 
with the {\it Spitzer Space Telescope}.  Infrared 
observations measure the infrared excess, which can 
be converted to a mass-loss rates assuming an expansion 
velocity for the circumstellar material.  So far all of the 
mass-loss rates have been estimated using the assumptions 
that the expansion velocity is independent of the 
metallicity.  The results presented here show that this 
assumption needs to be reconsidered.

We have recently obtained spectra of some of the present
sample with the Infrared Spectrograph on {\it Spitzer}.
Combining our CO observations and these spectra together 
and comparing them to {\it Spitzer} spectra from carbon 
stars in other galaxies in the Local Group will allow us 
to quantitatively study the mass-loss from carbon-rich AGB 
stars in the Local Group and its dependence on metallicity.  
This will be the subject of a forthcoming paper.



\section*{Acknowledgments}
EL wishes to thank the JAC staff for their great help 
carrying out this program, and Rodrigo Ibata for useful 
discussions about membership of stars to the Sgr stream. 
we thank the referee C. Loup for her useful comments that helped improving the quality of the paper.

\label{lastpage}

\end{document}